\documentclass[english,aps,prstper,reprint,showpacs,titlepage,longbibliography]{revtex4-2}

\usepackage[T1]{fontenc}
\usepackage[latin9]{inputenc}
\usepackage{geometry}
\usepackage{graphicx}
\usepackage[above,below]{placeins}
\usepackage{times}

\newcommand{\goeslike}{``goes like''}

\begin{document}

\begin{titlepage}

  \title{Exploring student facility with ``goes like'' reasoning in introductory physics}

  \author{Charlotte Zimmerman}
  \author{Alexis Olsho}
  \affiliation{Department of Physics, University of Washington, Seattle, WA, 98103} 
  \author{Andrew Boudreaux}
  \affiliation{Deprtment of Physics, Western Washington University, Bellingham, WA, 98225}
  \author{Trevor Smith}
  \affiliation{Department of Physics, Rowan University, Glassboro, NJ, 08028}
  \author{Philip Eaton}
  \affiliation{School of Natural Sciences and Mathematics, Stockton University, Galloway, NJ 08205}
  \author{Suzanne White Brahmia}
  \affiliation{Department of Physics, University of Washington, Seattle, WA, 98103} 

  \begin{abstract}
    Covariational reasoning---reasoning about how changes in one quantity relate to changes in another quantity---has been examined extensively in mathematics education research. Little research has been done, however, on covariational reasoning in introductory physics contexts. We explore one aspect of covariational reasoning: ``goes like'' reasoning. ``Goes like'' reasoning refers to ways physicists relate two quantities through a simplified function. For example, physicists often say that ``the electric field goes like one over r squared.'' While this reasoning mode is used regularly by physicists and physics instructors, how students make sense of and use it remains unclear. We present evidence from reasoning inventory items which indicate that many students are sense making with tools from prior math instruction, that could be developed into expert ``goes like'' thinking with direct instruction. Recommendations for further work in characterizing student sense making as a foundation for future development of instruction are made. 
  \clearpage
  \end{abstract}
    
  \maketitle
\end{titlepage}

\section{Introduction}
A perhaps unexpected byproduct of the COVID-19 pandemic is renewed clarity on how challenging it is for many to conceptualize the exponential function. This is certainly not novel; Albert Bartlett famously stated ``The greatest shortcoming of the human race is our inability to understand the exponential function''\cite{Bartlett1969ArithmeticEnergy}. This has become a public issue in the face of the coronavirus epidemic. Headlines such as ``What Does Exponential Growth Mean in the Context of COVID-19?,''\cite{washpost} ``The Exponential Power of Now,''\cite{nytimes} and ``Is Poor Math Literacy Making It Harder For People To Understand COVID-19 Coronavirus?''\cite{forbes} have put conceptualization of function on the national stage. 

It is evident that quantitative literacy---the set of skills that support the use of mathematics to describe and understand the world---is important, and lacking, in the United States today. Quantitative literacy has many facets, including reasoning about signed quantities, proportional reasoning and \textit{covariational reasoning}---conceptualizing change in one quantity with respect to change in another quantity \cite{brahmia2016a, Carlson2002, White2020neg}. Introductory physics, a broadly-required college course with a focus on quantifying and modeling nature, is an excellent place to address this need. 

Proportional reasoning---reasoning about ratio as a quantity---has been identified as critical for success in physics by physics educators and in Physics Education Research (PER). Early PER, confounded by student difficulties using elementary mathematics in physics contexts, focused on identifying specific reasoning difficulties such as the tendency to use additive, rather than multiplicative, strategies and the tendency of physics students to manipulate mathematical formalism without understanding the physical meaning of the associated quantities and operations \cite{Karplus1970IntellecturalSurvey, Arons1976CultivatingCourse, Arons1990ATeaching}. By the early 1980's, studies in PER had begun to systematically document and extend this body of work by using individual demonstration interviews to explore student understanding of velocity as the ratio $\Delta x / \Delta t$ and acceleration as the ratio $\Delta v / \Delta t$ \cite{Trowbridge1980InvestigationDimension, Trowbridge1981InvestigationDimension, Karplus1983EarlyProblems}. More recent work has examined the relationship between basic reasoning ability, including proportional reasoning, and the learning of physics content \cite{Coletta2005InterpretingAbility}. 


Work on the role and challenge of proportional reasoning in physics contexts has included attention to scaling and functional reasoning. Arons points out, for example, that few students ``have formed any conception of the basic function relation between area and linear dimensions,'' and that consequently, most students are ``unaware that all areas vary as the square of the length factor'' \cite{Arons1990ATeaching}. We build on this body of work by integrating the language of covariational reasoning established by Research in Undergraduate Mathematics Education (RUME) community \cite{Thompson1994rate, saldanha1998, moore2013, Carlson2002, oehrtman2008, Carlson2010precalculus}. Covariation encompasses all functions that relate two or more quantities and considers multiple ways that one can think about those relationships. For example, one can consider discrete covariation (if the radius is doubled, what happens to the electric field at a point?), or continuous covariation (how does the field change smoothly as the radius is increased?) \cite{Carlson2002}. We suggest that proportional reasoning is a subset of covariational reasoning, focused specifically on linear relationships and using ratios that have meaning as a single entity (such as velocity and acceleration). 

Physics educators regularly identify ``thinking like a physicist'' as a goal of introductory physics. In a 2019 study of the ways in which experts use covariational reasoning to solve introductory physics problems by Zimmerman, Olsho, Boudreaux, Loverude, and White Brahmia, 
it was noted that physics experts use functional reasoning by employing the ``$\propto$'' symbol or phrases like ``goes like'' to illustrate relationships in statements like $\text{Area} \propto  r^2$, Force goes like $1/r^2$, etc. \cite{ZImmerman2019perc}  This kind of ``goes like'' expert thinking is used to represent a wide variety of simplified relationships between quantities, and is a desired outcome of introductory physics instruction. In his work on proportional reasoning, Arons asserts that the capacity for scaling and functional reasoning will not necessarily develop spontaneously \cite{Arons1990ATeaching}. Indeed, the need for curricular intervention is evident from the current literature. What is less clear is what resources and emergent abilities students \textit{do have} regarding quantitative literacy prior to physics instruction, and what educators can do to build upon these skills to develop quantitative reasoning in their students.

This paper describes a study of students' covariational reasoning in physics contexts. It contributes to the work in mathematics education, as well as to closing a gap in PER, where it has been shown that reasoning in physics contexts is different from reasoning in purely mathematical ones \cite{redishkuo, Pollock2007}. We focus on one expert-like facet of physics covariational reasoning: ``goes like'' reasoning \cite{ZImmerman2019perc}. We will present some of the ways an expert might use this kind of thinking, and some preliminary results that suggest while introductory students do not have strong facility with physics ``goes like'' reasoning, and their conceptualization of ``goes like'' is not improving over a year long sequence in introductory physics, they do have some productive resources and emergent abilities from prior math courses that can be met and built upon with direct instruction to develop physics covariational reasoning skills.  Recommendations for future work and curricular interventions are made.

\section{Expert Reasoning About ``Goes Like''}
``Goes like'' reasoning refers to the simplified relationship between two changing quantities that illustrates the behavior of an evolving system. For example, consider a classic introductory physics problem: a ball thrown from a cliff. An expert might reason that if the ball's initial height is increased, the final speed of the ball will also increase. They might reason further that the final speed of the ball ``goes like'' the the square root of the height. Here, ``goes like'' reasoning allows the expert to focus on the functional form of the relationship between two changing quantities, and to ignore any constants or pre-factors. This focus on co-varying quantities in turn allows for efficient problem solving, as the expected behavior of the system can be quickly and clearly illustrated.

We note that expert use of \goeslike~reasoning relies on facility with the mathematical functions involved, as well as the experience and physics content knowledge that enable experts to relate physical phenomena to those functions. Zimmerman et. al 
found that physics graduate students have strong associations between certain routinely used physics quantities that allow them to make inferences about relationships between quantities in a given problem. This simplifies problems to those they can solve more efficiently, or to which they may already know the answer from experience \cite{ZImmerman2019perc}. Unlike novices, someone with substantial experience with physics is able to make claims such as ``This problem involves a potential, which goes like $1/r$'' or ``This looks like scattering, so I expect it to be an exponential.''

We don't claim that novices do not have some ''compiled relationships'' between mathematical functions and physical contexts. To the contrary, in our experience many students have strong associations between functions that model real world contexts, and we consider these to be resources for physics learning. Many of these associations evolve from prior math instruction and so are more suited to math contexts than to a physics course---for example, where experts may associate circular motion with sinusoidal curves, introductory physics students may more readily associate trigonometric functions with right triangles.


This led us to wonder what resources students in an introductory physics course are using to relate two quantities, and whether they include ``goes like'' reasoning. In particular, we asked: do students enter introductory physics with this skill already formed and ready to be applied from math courses? In addition, do their ``goes like'' reasoning skills improve after instruction in a physics class, where instruction typically takes the form of experts modeling their reasoning and discussing it in lecture? To answer these questions, we probed students' covariational reasoning using items from an inventory currently in development: the Physics Inventory for Quantitative Reasoning (PIQL) \cite{Smith2019rume, White2020neg}. 

\section{Assessing ``Goes Like'' Reasoning}
The PIQL measures fundamental aspects of mathematical reasoning that are ubiquitous in physics modeling, i.e. \textit{physics quantitative literacy (PQL)}. Development of this instrument began with items targeting proportional reasoning and reasoning about sign and signed quantities, and has since grown to include additional items related to covariational reasoning more broadly. During its development, the PIQL has been administered over several years at a large research university in the Pacific Northwest. The test is given at the start of each of the three quarter-long courses that form a year-long introductory physics sequence. Here we focus on student responses to two PIQL items that we believe would illicit ``goes like'' reasoning in experts: the Flag of Bhutan and Ferris Wheel \cite{boudreaux2015, Hobson2017}. Other aspects of quantification and PQL are also involved in these responses, but will not be discussed in this paper.

\subsection{Flag of Bhutan}
\begin{figure}
  \includegraphics[width=0.55\linewidth]{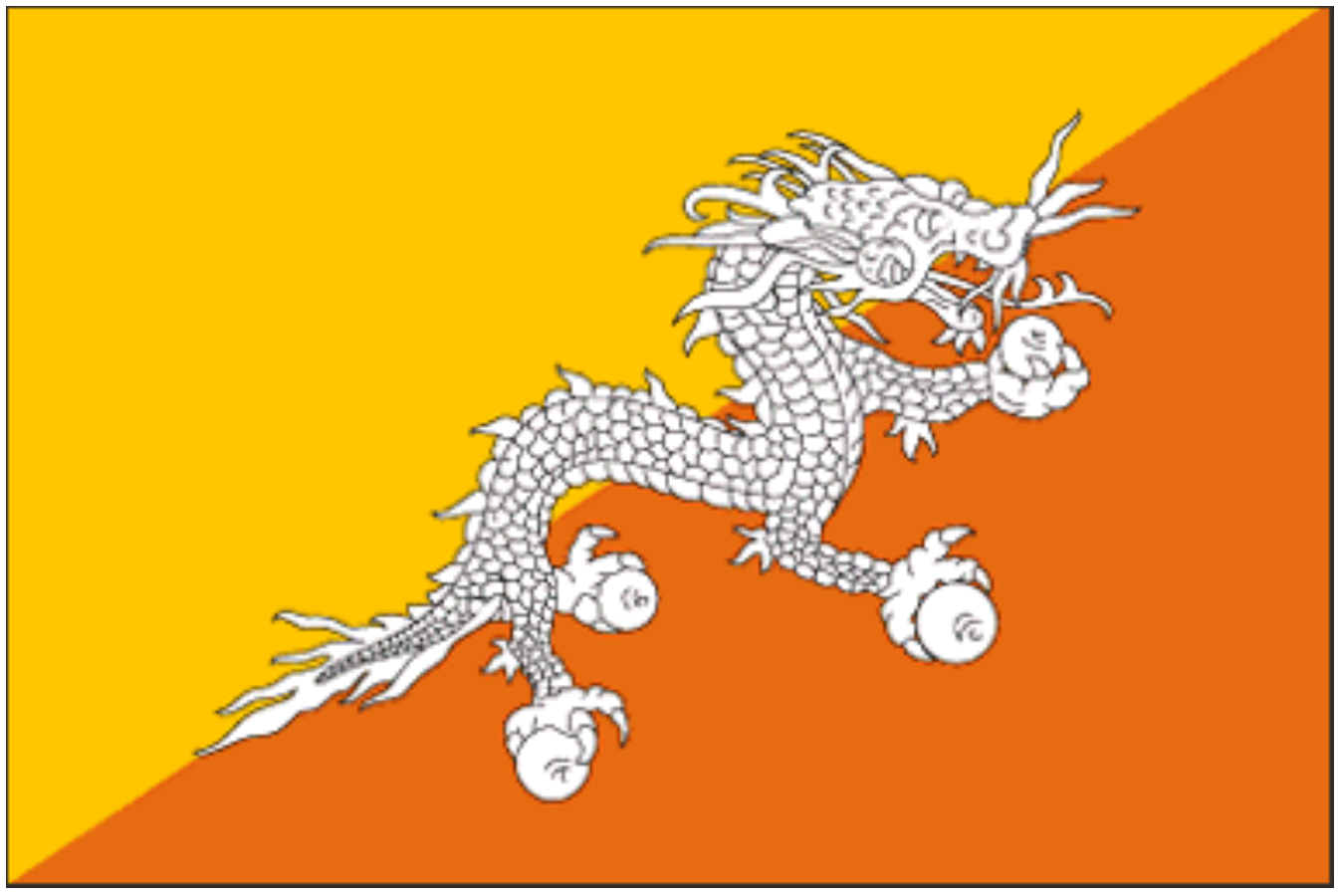}
  \caption{The Flag of Bhutan. The prompt associated with this image asks students to \textbf{select all of the following quantities} that are larger by a factor of 1.5 when the length and width of the flag are both increased by a factor of 1.5: (a) The distance around the edge of the flag, (b) the amount of cloth needed to make the flag, (c) the length of the curve forming the dragon's backbone, (d) the diagonal of the flag, and (e) none of these. Students are prompted to choose all answer choices that apply. We believe the correct answers to be (a), (c) and (d).}
  \label{fig:bhutan}
\end{figure}

In the Flag of Bhutan question, students are asked to determine what aspects of the flag would be larger by a factor of 1.5 if the length and the width were both increased by a factor of 1.5 (see Fig.\ \ref{fig:bhutan}). This item was originally designed as a scaling assessment to measure student facility with both linear and non-linear relationships, as some answer choices depend linearly on length and width (such as the length of the dragon's backbone, or the distance around the edge of the flag) and the answer choice ``the amount of cloth needed to make the flag'' depends on length times width \cite{boudreaux2015}. While scaling was considered a facet of proportional reasoning at the time, it was understood by the researchers that scaling with non-linear functions is notably different than scaling with linear relationships. We believe that this question can be re-examined in the context of discrete covariation and ``goes like'' thinking.

One of the challenges of the Flag of Bhutan item is that it is a multiple-choice/multiple-response (MCMR) question. Thereby, its score is low compared to other items on the PIQL because these items are scored dichotomously for comparison with other multiple-choice/single-response items \cite{Smith2019a}. However, the nature of the item does not fully account for the significantly low number of completely correct responses. Results suggest a majority of introductory students struggle to reason without a linear equation as only 26\% of students answer completely correctly, in contrast to instructors' expectations. Moreover, the percentage of students who answer this question completely correctly does not change significantly throughout the introductory sequence (25\% in the first quarter, 25\% in the second quarter, and 31\% in the third quarter) suggesting that this kind of reasoning does not improve. 


One of the benefits of an MCMR item is that we can learn more about student thinking by examining the partially correct answers that students chose. The most common reasons a student does not get the item completely correct are by not selecting either (c) or (d) in Figure~\ref{fig:bhutan}. These were not chosen by 55\% and 43\% of students respectively. Only 24\% of students do not choose (a). These results suggest that students do have facility with directly linear relationships, such as perimeter to length and width, but have difficulty with more complex functional relationships, such as $\sqrt{l^2 + w^2}$, or those that do not have a known functional relationship, such as the dragon's backbone, even if the result is linear. Using tetrachoric correlation analysis, we found that students considered (c) and (d) together (either choosing both or declining to choose both) 66\% of the time. This suggests that even while the diagonal of the flag can be described by a geometric function and the backbone cannot, the majority of students are able to realize that they have the same dependence on length and width. However, these results do not improve over the course of instruction, suggesting that these early signs of ``goes like'' reasoning might not be nurtured over the course of instruction to develop students' discrete covariational reasoning in the context of scaling. 



\subsection{Ferris Wheel}
\begin{figure}
  \includegraphics[width=0.5\linewidth]{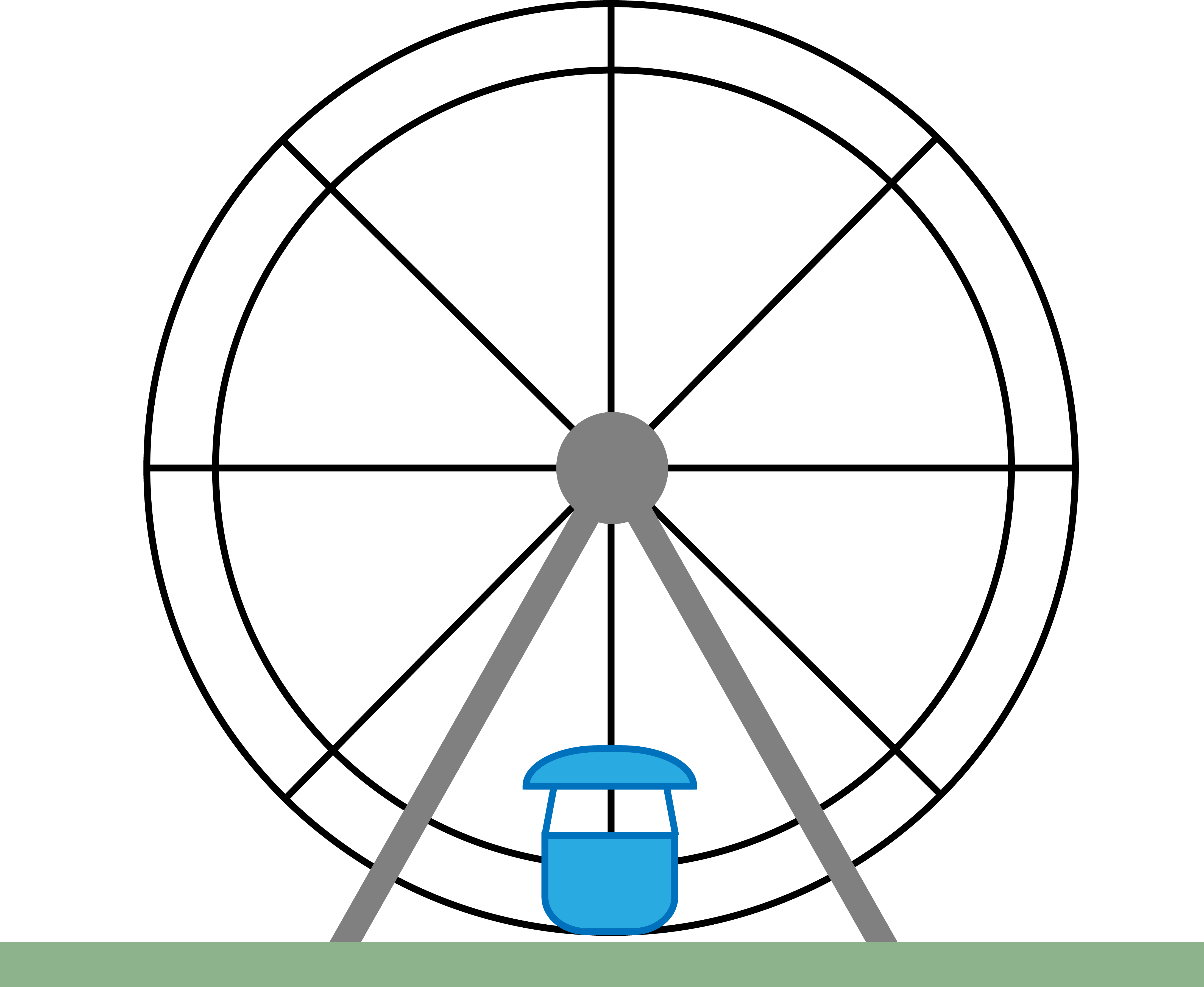}
  \caption{Ferris Wheel. The prompt associated with this image asks students to identify which expression correctly identifies how $h$, the height of the cart, directly changes with $s$, the distance traveled by the rider, where the radius of the Ferris wheel is given by $R_0$: (a) $h(s) = \sqrt{s^2 + R_0^2}$, (b) $h(s) = R_0 \exp(s/R_0)$, (c) $h(s) = R_0 - R_0 \cos(s/R_0)$, (d) $h(s) = s^2 / (2\pi R_0)$}
  \label{fig:ferris}
 \end{figure}

Ferris Wheel asks students to choose an equation that represents how the height of a Ferris wheel cart changes as a function of the total distance it has traveled (see Fig.\ \ref{fig:ferris}). This question was inspired by a Hobson and Moore study, and the distractors were developed based on results from the Zimmerman et al. study, and introductory student interviews \cite{Hobson2017, ZImmerman2019perc}. Experts were given an animated version of the image in Figure ~\ref{fig:ferris} in which the cart rotates with the Ferris wheel, and asked to produce a graph that relates the total distance traveled by the cart and the height of the cart \cite{Hobson2017, ZImmerman2019perc}. It was observed that the experts used time as a proxy for total distance, noticing that both quantities described the evolution of the system. They then demonstrated ``goes like'' reasoning by making strong associations between the circular motion presented in the animation and trigonometric functions: ``the height goes like a trig function.'' The authors refer to these connections between quantities as \textit{compiled relationships} \cite{ZImmerman2019perc}. In developing Ferris Wheel for the PIQL, we were interested to see if students also held compiled relationships, and if they were able to use ``goes like'' reasoning to solve the problem. When reformating the item as multiple choice, we tried to choose distractors that represented other potential compiled relationships based on geometric shapes including the Pythagorean theorem, which students associate with triangles in interviews and in open-ended versions of other PIQL items, and an expression containing the circumference, which students associate with circles. 

Ferris Wheel was administered as part of the PIQL, and validation interviews were performed at another public university in the Pacific Northwest. We do not claim that these two institutions represent identical populations; they often have slightly different average scores on PIQL assessment items. Indeed, while a majority of students that took the PIQL at the large research university answered the question correctly (58\%), fewer than half of those interviewed chose the correct answer. However, the interviews do provide a broad look into how some students are making sense of the problem. Based on the percentage of students who get the assessment items completely correct, this item appears to be considerably less challenging than Flag of Bhutan. However, it  is not an MCMR question, so it cannot be compared directly \cite{Smith2019a}.

As before, we can explore what students may be thinking by examining their incorrect answer choices. The most common incorrect choices were (d) and (a) from Figure~\ref{fig:ferris}, with an answer rate 24.5\% and 14.7\% respectively that does not change significantly across the three courses. These results suggest that the ``circumference-like'' distractor and the ``Pythagorean-like'' distractor are appealing to a significant fraction of the entire student population. We interpret these answer choices as unrefined ``goes like'' tendencies---these are functions that are familiar to students from recent math classes, and have been fruitful in past experiences reasoning about circles and triangles.  Some students may not have readily accessible resources of ``goes like'' reasoning in physics contexts, or a compiled relationship between circular-motion and trigonometry as demonstrated by physics experts, even though they are making sense with the tools they have. 
 
Validation interviews can provide some details into what compiled relationships students have formed, and how they might be using them along with ``goes like'' reasoning to solve the problem. It was found that nearly all students interviewed were highly invested in answer choice (d), citing it as familiar. They often noted that it contains the expression for circumference of a circle, which most of those interviewed readily associated with the total distance traveled: ``I'd say (d) because its the only one that has $2\pi R$ in there, which is the, essentially, the circumference formula.'' Indeed, nearly all students interviewed began by defining the total distance traveled by the circumference, and many returned to this definition throughout their problem solving process. While experts may realize that focusing on circumference is not a productive method of solving this problem, we recognize this as a form of quantitative reasoning---students demonstrate a strong compiled relationship between distance and circumference. The key difference is that experts are able to use distance as a quantity that describes the evolution of the system, while students are connecting total distance traveled (a quantity that changes in time) with the circumference of one revolution (a quantity that is fixed in this problem). Because the students interviewed didn't spontaneously consider the total distance \textit{as it is changing}, they didn't demonstrate facility with expert-like ``goes like'' reasoning. They did not reach the point in the problem where they could choose an expression for the height as a function of total distance with confidence. There was only one student who articulated that the total distance is a changing quantity stating it represented ``how much of the circumference [the rider] has traveled,'' but in this student's case that line of reasoning was still used to support his selection of answer choice (d).

Interest in answer choice (a) was centered around reasoning with triangles, and although none of the students interviewed chose (a) as their final answer, many grappled with its meaning. Every student interviewed verbally labelled option (a) as ``Pythagorean,'' and many students drew an accompanying triangle, demonstrating a strong compiled relationship with the expression itself and triangular geometry. Some recognized right away that the Pythagorean approach would not work, one stating, ``(A) is the Pythagorean theorem, but that doesn't make sense because that's linear distance.'' Here, we interpret this as the student recognizing that Pythagorean theorem uses linear distances, and the total distance traveled is not linear. Another student interviewed debated about the correctness of (a), stating, ``This is like the Pythagorean theorem\ldots if we do it like this, [the student draws a triangle with the hypotenuse representing total distance] I guess you could estimate [the total distance] as being a straight line.'' Both of these students did not draw the triangle a physics expert might expect (with the radius as the hypotenuse), and most were uncertain about the expression presented because they had difficulty making sense of which quantities the sides of the triangle they drew were representing. However, their statements  demonstrate sense making about the expression and its connection to right triangles, which we consider to be productive. 

When evaluating answer choice (c) that uses a trigonometric expression, students continued to puzzle over how to draw the appropriate triangle: ``cosine gives me $s$ over $R_0$\ldots so they're saying the radius is the hypotenuse. How can that be?'' Only one student interviewed made direct reference to the unit circle and was able to quickly recognize that ``$\theta$ is equal to arc length over the radius,'' and that ``the radius should be the hypotenuse because the radius is the one thing that is measured throughout the circle,'' but then this student was drawn to the familiarity of (d) and eventually uses point by point analysis to choose her answer. These patterns suggest that the students interviewed have strong procedural facility with a geometric approach to Pythagorean theorem, but not conceptual understanding about how it connects to circles. This gap in understanding between trigonometry learned and how it is applied in physics was typical in the interviews. It is notable that while students may comfortably reason about trigonometry in the contexts of triangles and circles, many students may not understand how that reasoning is used in physics contexts.

Those that answered correctly in interviews often determined their answer by plugging in points. Uniformly this strategy was approached as a last effort, suggesting that students don't rely on other ways of making sense of the answer choices and may consider plugging in numbers to be an expert problem solving strategy. Typically, students using this method were choosing between option (c) and (d), however in one case the student tried all possible answer choices. In particular, students that did pick points to solve the problem choose physically significant points, for example, the bottom and top of the Ferris Wheel where the height is at a minimum or maximum. This kind of problem solving---specifically choosing physically relevant points to better understand the behavior of the system---has been identified as an expert-like behavior in previous studies \cite{ZImmerman2019perc}.

\section{Conclusions}
Ferris Wheel and Flag of Bhutan demonstrate that while students have difficulty with physics ``goes like'' reasoning, they illustrate skills that could be used to develop physics covariational reasoning with direct instruction. Responses to Flag of Bhutan show that students have strong ``goes like'' reasoning about linear relationships that could be developed into ``goes like'' reasoning about non-linear relationships. Responses to Ferris Wheel demonstrate that students have strong compiled relationships regarding right triangles and the Pythagorean theorem, and circles and circumference, that could be developed into compiled relationships between circular motion and trigonometric functions. As covariational reasoning is integral to conceptualizing physics models, we recommend instructors consider including direct and explicit instruction on relating quantities beyond demonstration in their own teaching. Additional studies are needed to better understand what kinds of covariational reasoning and compiled relationships students have coming into introductory physics and are forming over the course of instruction. Currently, appropriate curricular materials do not exist and need to be developed.

\acknowledgments{We'd like to thank Steve Kanim for his part in developing Flag of Bhutan. This work is supported by the University of Washington
and the National Science Foundation under grants DUE-1832836 and DGE-1762114.}


\begin{thebibliography}{99}
  
  \bibitem{Bartlett1969ArithmeticEnergy} A. Bartlett, Arithmetic, Population and Energy (1969).
	\bibitem{washpost} M. Cappucci, What does exponential growth mean in the context of covid-19?, The Washington Post (March 27, 2020).
	\bibitem{nytimes} S. Roberts, The Exponential Power of Now, The New York Times (March 13, 2020).
	\bibitem{forbes} M. Sheperd, Is Poor Math Literacy Making It Harder for People to Understand COVID-19 Coronavirus?, Forbes (March 23, 2020).
	\bibitem{brahmia2016a} S. Brahmia, A. Boudreaux, and S. E. Kanim, Obstacles to Mathematization in Introductory Physics, ArXiv e-prints (2016).
	\bibitem{Carlson2002} M. Carlson, S. Jacobs, E. Coe, S. Larson, and E. Hsu, Applying Covariational Reasoning While Modeling Dynamic Events: A Framework and a Study, Journal for Research in Mathematics Education \textbf{33}, 352 (2002).
	\bibitem{White2020neg} S. White Brahmia, A. Olsho, T. I. Smith, and A. Boudreaux, Framework for the natures of negativity in introductory physics, Phys. Rev. Phys. Educ. Res. \textbf{16} 010120 (2020).
	\bibitem{Karplus1970IntellecturalSurvey} R. Karplus and R. W. Peterson, Intellectural development beyond elementary school II: Ratio, a survey, School Sci. Math. \textbf{70}, 813 (1970).
	\bibitem{Arons1976CultivatingCourse} A. B. Arons, Cultivating the capactiy for formal reasoning: Objects and procedures in an introductory physical science course, American Journal of Physics \textbf{44}, 834 (1976).
	\bibitem{Arons1990ATeaching} A. B. Arons, \textit{A Guide to Introductory Physics Teaching} (Wiley, New York, NY, 1990) p. 342.
	\bibitem{Trowbridge1980InvestigationDimension} D. Trowbridge and L. C. McDermott, Investigation of student understanding of the concept of velocity in one dimension, American Journal of Physics \textbf{49}, 1020 (1980).
	\bibitem{Trowbridge1981InvestigationDimension} D. Trowbridge and L. C. McDermott, Investigation of student understanding of the concept of acceleration in one dimension, American Journal of Physics \textbf{48}, 242 (1981).
	\bibitem{Karplus1983EarlyProblems} R. Karplus, E. K. Stage, and S. Pulos, Early adolescents' proportional reasoning on `rate' problems, Educational Studies in Mathematics \textbf{13}, 219 (1983).
	\bibitem{Coletta2005InterpretingAbility} V. Coletta and J. Philips, Interpreting FCI scores: Normalized gain, preinstruction scores, and scientific reasoning ability, American Journal of Physics \textbf{73}, 1172 (2005).
	\bibitem{Thompson1994rate} P. W. Thompson, Images of rate and operational understanding of the fundamental theorem of calculus, Educational Studies in Mathematics 10.1007/BF01273664 (1994).
	\bibitem{saldanha1998} L. A. Saldanha and P. W. Thompson, Re-thinking co-variation from a quantitative perspective: Simultaneous continuous variation, in \textit{Annual Meeting of the Psychology of Mathematics Education --- North America}, Vol. 1, edited by S. B. Berenson and W. N. Coulombe (North Carolina State University, Raleigh, NC, 1998) pp. 298-304.
	\bibitem{moore2013} K. C. Moore, T. Paoletti, and S. Musgrave, Covariational reasoning and invariance among coordinate systems, Journal of Mathematical Behavior \textbf{32}, 461 (2013).
	\bibitem{oehrtman2008} M.  Oehrtman,  M.  Carlson,  and  P.  W.  Thompson,  Foundational  Reasoning  Abilities  that  Promote  Coherence  in  Students' Function Understanding, in \textit{Making the Connection: Research and Teaching in Undergraduate Mathematics Education}, Vol. 73 (Mathematical Association of America, 2008) 1st ed., pp. 27-42.
	\bibitem{Carlson2010precalculus} M. Carlson, M. Oehrtman, and N. Engelke, The precalculus concept assessment: A tool for assessing students' reasoning abilities and understandings, Cognition and Instruction \textbf{28}, 113 (2010).
	\bibitem{ZImmerman2019perc} C. Zimmerman, A. Olsho, S. White Brahmia, M. Loverude, A. Boudreaux, and T. I. Smith, Toward understanding and characterizeing expert physics covariational reasoning, in \textit{Physics Education Research Conference 2019}, PER Conference (Provo, UT, 2019).
	\bibitem{redishkuo} E. F. Redish and E. Kuo, Language of physics, language of math: Disciplinary culture and dynamic epistemology, Science \& Education \textbf{24}, 561 (2015).
	\bibitem{Pollock2007} E. B. Pollock, J. R. Thompson, and D. B. Mountcastle, Student understanding of the physics and mathematics of process variables in P-V diagrams in \textit{2007 Physics Education Research Conference, AIP Conference Proceedings}, Vol. 951, edited by L. Hsu, C. Henderson, and L. McCullough, American Association of Physics Teachers (American Institute of Physics, Melville, NY 2007) pp. 168-171.
	\bibitem{Smith2019rume} T. I. Smith, S. White Brahmia, A. Olsho, and A. Boudreaux, Developing a reasoning inventory for measuring physics quantitative literacy, in \textit{Proceedings of the 22nd Annual Conference on Research in Undergraduate Mathematics Education}, edited by A. Weinberg, D. Moore-Russo, H. Soto, and M. Wawro (Oklahoma City, OK, 2019) pp. 1181-1182.
	\bibitem{boudreaux2015} A. Boudreaux, S. Kanim and S. Brahmia, Student facility with ratio and proportion: Mapping the reasoning space in introductory physics, arXiv preprint arXiv:1511.08960 (2015).
	\bibitem{Hobson2017} N. L. F. Hobson and K. C. Moore, Exploring experts' covariational reasoning, in \textit{20th Annual Conference on Research in Undergraduate Mathematics Education} (Moore \& Thompson, 2017) pp. 664-672.
	\bibitem{Smith2019a} T. I. Smith, K. J. Louis, B. J. Ricci, and N. Bendjilali, Quantitatively ranking incorrect responses to multiple-choice questions using item response theory, (under review) preprint available arXiv:1906.00521 (2019).
	
\end{thebibliography}
\end{document}